\documentclass[journal]{IEEEtran}
\usepackage[utf8]{inputenc} %unicode support
\usepackage{graphicx}
\usepackage{subfigure}
\usepackage{amsmath}
\usepackage{multirow}
\usepackage[varg]{newtxmath}
\usepackage{booktabs}
\usepackage{amsmath}

\usepackage[psamsfonts]{amssymb}
\usepackage{amsxtra}
\usepackage{threeparttable}
\usepackage{romannum}
\usepackage{stfloats}
\usepackage{cite}
\usepackage{color, xcolor}
\usepackage{threeparttable}
\usepackage{ulem}
\usepackage{algorithmic}
\usepackage{algorithm}

\ifCLASSINFOpdf
\else
\fi
\begin{document}

\title{SSDPT: Self-Supervised Dual-Path Transformer \\for Anomalous Sound Detection in Machine Condition Monitoring}

%%%%%%%%%%%%%%%%%%%%%%%%%%%%%%%%%%%%%%%%%%%%%%
%%                                          %%
%% Enter the authors here                   %%
%%                                          %%
%% Specify information, if available,       %%
%% in the form:                             %%
%%   <key>={<id1>,<id2>}                    %%
%%   <key>=                                 %%
%% Comment or delete the keys which are     %%
%% not used. Repeat \author command as much %%
%% as required.                             %%
%%                                          %%
%%%%%%%%%%%%%%%%%%%%%%%%%%%%%%%%%%%%%%%%%%%%%%

\author{Jisheng Bai,~\IEEEmembership{Student~Member,~IEEE,}
	Jianfeng Chen,~\IEEEmembership{Senior Member,~IEEE,}
	Mou Wang,~\IEEEmembership{Student~Member,~IEEE}
	Muhammad Saad Ayub,
	and Qingli Yan,
	%and~Xiao-Lei Zhang,~\IEEEmembership{Senior Member,~IEEE}
	% <-this % stops a space
	\thanks{This work is supported by the National Natural Science Foundation of China under Grant No.62071383 and the Key research and development plan of Shaanxi Province (2021NY-036). \textit{(Corresponding author: Jianfeng Chen.)} }
    \thanks{J. Bai, J. Chen, M. Wang and M. S. Ayub are with the Joint Laboratory of Environmental Sound Sensing, School of Marine Science and Technology, Northwestern Polytechnical University, Xi’an, China, and also with the LianFeng Acoustic Technologies Co., Ltd. Xi'an, China. (e-mail:  baijs@mail.nwpu.edu.cn; chenjf@nwpu.edu.cn; wangmou21@mail.nwpu.edu.cn; msaadayub@mail.nwpu.edu.cn)}
    \thanks{Qingli Yan is with the Xi’an University of Posts and Telecommunications, Xi’an, China. (e-mail: yql@xupt.edu.cn)
	}% <-this % stops a space  ; xiaolei.zhang@nwpu.edu.cn
	% <-this % stops a space
}

%%%%%%%%%%%%%%%%%%%%%%%%%%%%%%%%%%%%%%%%%%%%%%
%%                                          %%
%% Enter the authors' addresses here        %%
%%                                          %%
%% Repeat \address commands as much as      %%
%% required.                                %%
%%                                          %%
%%%%%%%%%%%%%%%%%%%%%%%%%%%%%%%%%%%%%%%%%%%%%%

%%%%%%%%%%%%%%%%%%%%%%%%%%%%%%%%%%%%%%%%%%%%%%
%%                                          %%
%% Enter short notes here                   %%
%%                                          %%
%% Short notes will be after addresses      %%
%% on first page.                           %%
%%                                          %%
%%%%%%%%%%%%%%%%%%%%%%%%%%%%%%%%%%%%%%%%%%%%%%
\maketitle
\pagenumbering{arabic}
%\end{fmbox}% comment this for two column layout

%%%%%%%%%%%%%%%%%%%%%%%%%%%%%%%%%%%%%%%%%%%%%%
%%                                          %%
%% The Abstract begins here                 %%
%%                                          %%
%% Please refer to the Instructions for     %%
%% authors on http://www.biomedcentral.com  %%
%% and include the section headings         %%
%% accordingly for your article type.       %%
%%                                          %%
%%%%%%%%%%%%%%%%%%%%%%%%%%%%%%%%%%%%%%%%%%%%%%

\begin{abstract} % abstract
Anomalous sound detection for machine condition monitoring has great potential in the development of Industry 4.0. 
However, these anomalous sounds of machines are usually unavailable in normal conditions.
Therefore, the models employed have to learn acoustic representations with normal sounds for training, and detect anomalous sounds while testing.
In this article, we propose a self-supervised dual-path Transformer (SSDPT) network to detect anomalous sounds in machine monitoring.
The SSDPT network splits the acoustic features into segments and employs several DPT blocks for time and frequency modeling.
DPT blocks use attention modules to alternately model the interactive information about the frequency and temporal components of the segmented acoustic features.
To address the problem of lack of anomalous sound, we adopt a self-supervised learning approach to train the network with normal sound. 
Specifically, this approach randomly masks and reconstructs the acoustic features, and jointly classifies machine identity information to improve the performance of anomalous sound detection. 
We evaluated our method on the DCASE2021 task2 dataset. 
The experimental results show that the SSDPT network achieves a significant increase in the harmonic mean AUC score, in comparison to present state-of-the-art methods of anomalous sound detection.
	
%\parttitle{First part title} %if any
%Text for this section.

%\parttitle{Second part title} %if any
%Text for this section.
\end{abstract}

%%%%%%%%%%%%%%%%%%%%%%%%%%%%%%%%%%%%%%%%%%%%%%
%%                                          %%
%% The keywords begin here                  %%
%%                                          %%
%% Put each keyword in separate \kwd{}.     %%
%%                                          %%
%%%%%%%%%%%%%%%%%%%%%%%%%%%%%%%%%%%%%%%%%%%%%%


% Self-Supervised Vision-Based Detection of the Active Speaker as Support for Socially Aware Language Acquisition
%Anomalous Behaviors Detection in Moving Crowds Based on a Weighted Convolutional Autoencoder-Long Short-Term Memory Network

% MSC classifications codes, if any
%\begin{keyword}[class=AMS]
%\kwd[Primary ]{}
%\kwd{}
%\kwd[; secondary ]{}
%\end{keyword}

\IEEEpeerreviewmaketitle

%%%%%%%%%%%%%%%%%%%%%%%%%%%%%%%%%%%%%%%%%%%%%%
%%                                          %%
%% The Main Body begins here                %%
%%                                          %%
%% Please refer to the instructions for     %%
%% authors on:                              %%
%% http://www.biomedcentral.com/info/authors%%
%% and include the section headings         %%
%% accordingly for your article type.       %%
%%                                          %%
%% See the Results and Discussion section   %%
%% for details on how to create sub-sections%%
%%                                          %%
%% use \cite{...} to cite references        %%
%%  \cite{koon} and                         %%
%%  \cite{oreg,khar,zvai,xjon,schn,pond}    %%
%%  \nocite{smith,marg,hunn,advi,koha,mouse}%%
%%                                          %%
%%%%%%%%%%%%%%%%%%%%%%%%%%%%%%%%%%%%%%%%%%%%%%

%%%%%%%%%%%%%%%%%%%%%%%%% start of article main body
% <put your article body there>

%%%%%%%%%%%%%%%%
%% Background %%
%%
%\section*{Content}
%Text and results for this section, as per the individual journal's instructions for authors. %\cite{koon,oreg,khar,zvai,xjon,schn,pond,smith,marg,hunn,advi,koha,mouse}

\section{Introduction}
\label{sec:intro}

Machine condition monitoring (MCM) plays an important role in factory automation \cite{dinardo2018smart}.
Sound monitoring has many advantages, e.g., a pump suffering from a small leakage might not be inspected visually, but it can be detected acoustically by observing distinct audio patterns \cite{yamashita2006inspection}. 
The early detection of mechanical anomalies with a reliable acoustic system can prevent problems and reduce the cost of surveillance \cite{Sonntag2017}.
Anomalous sound detection (ASD) for MCM has been widely used in several applications to detect anomalous cases before causing damage \cite{espinosa2021click, nunes2021anomalous}. 

General ASD systems usually use machine learning methods to automatically detect anomalous sound, and these systems can be further categorized into two classes, i.e., supervised and unsupervised learning-based ASD systems \cite{koizumi2018unsupervised, marchi2015novel, ntalampiras2011probabilistic}.
In supervised methods, normal and abnormal sounds are available and annotated in advance \cite{lim2017rare}. 
However, supervised learning-based ASD systems face several challenges.
The most challenging issue is that anomalous sounds are rare and even unavailable.
It is expensive and unfeasible to break or impair the machines to collect anomalous sound samples, and we can not pre-define the anomalies if they are not observed.
Instead, it is possible to collect adequate normal sounds, so ASD systems use only employ sounds to train acoustic models.

ASD systems usually adopt unsupervised learning methods because only normal sounds are available in the training stage \cite{mnasri2022anomalous}. 
Autoencoders (AEs) are a typical type of unsupervised learning algorithm, which outperforms the previous methods for ASD \cite{Bai2020, Giri2020, an2015variational}. 
AEs usually transform the input data into latent features and then reconstruct them by minimizing the error between the reconstructed data and original data \cite{thill2021temporal}.
The characteristics of normal sounds can be properly represented by the latent features if the reconstruction error is very small.
To estimate the state of a target sound, an ASD system will calculate the anomaly score according to the reconstruction loss \cite{koizumi2018unsupervised, giri2020unsupervised}.
If the anomaly score surpasses a given threshold, the target sound is anomalous, and vice versa \cite{ruff2021unifying}. 

In recent years, self-supervised learning (SSL) has attracted much attention in many pattern recognition fields \cite{georgescu2021anomaly, li2021cutpaste, stefanov2019self}.
SSL focuses on a well-defined pretext task and simultaneously introduces additional tasks, called auxiliary tasks, to learn better latent representations \cite{huang2022efficient, kerzel2020enhancing}.
In SSL, the supervised information is generated from the training data, making it possible to learn useful representation without large labelled data. 
SSL has also been proposed to improve the performance of unsupervised audio pattern recognition \cite{tripathi2021self}. 
An SSL framework has been proposed for both speech and audio tasks, where the spectrogram patches are masked during training, forcing the model to learn both the temporal and frequency structure \cite{gong2021ssast}.
For ASD, robust representations of the normal sounds can be learned using the supervised information derived from these sounds.
In \cite{dohi2021flow}, the authors present a self-supervised ASD system, which introduces an auxiliary task using the IDs of machines for training a classifier, and this classifier will then predict a wrong machine ID if the target sound is anomalous.

The early machine learning-based ASD systems focused on conventional models, such as Gaussian mixture models \cite{ntalampiras2011probabilistic} or support vector machines \cite{foggia2015audio}. 
Recently, deep learning-based approaches have shown a great ability to extract deep representations and have been introduced in many ASD systems.
Convolutional neural networks (CNNs) can extract local invariant features, model the local time-frequency information of acoustic features, and perform better than AEs for ASD \cite{wang2019hybrid, bai2022multimodal, Primus2020}. 
However, CNNs face difficulty in modeling data with long sequences, so recurrent neural networks (RNNs) and Transformer are proposed for catching long temporal dependencies in ASD tasks.
While RNNs can not process information in parallel, Transformers have been proposed to address this problem.
The multi-head self-attention (MHSA) mechanism is able to capture global dependencies and process information efficiently \cite{vaswani2017attention}.
Transformer-based architectures have achieved state-of-the-art performance in computer vision, natural language processing, speech processing, and environmental sound recognition tasks \cite{dosovitskiy2020image, chen2020dual}.
Nevertheless, Transformer-based networks usually model on time, and neglect the frequency interrelationships among acoustic features. 

In this paper, we develop a self-supervised dual-path Transformer (SSDPT) network for ASD in MCM. 
The proposed SSDPT first takes the acoustic features as input and divides them into overlapped short segments.
Then the short segments are fed to the dual-path Transformer (DPT) for fine-grained modeling on two dimensions.
In each DPT block, we use Transformer encoders to alternately model the acoustic features on time and frequency dimensions.
To address the problem of availability of normal sounds only for training, we propose to train the DPT with two SSL strategies.
One of the strategies utilizes the metadata information (the IDs assigned to the machines) for classification. 
The other strategy is designed to learn better latent representations by randomly masking the areas of the acoustic features and reconstructing them.
For testing, the trained SSDPT network will output an anomaly score of a target sound to determine if it is anomalous. 
The anomaly score is comprised of two parts, the first part is the reconstruction loss between reconstruction features and original features, and the second part is the averaged negative logarithm of predicted probabilities for the correct machine ID.
The main contributions of this paper can be summarized as follows:
\begin{itemize}
	\item We propose a dual-path Transformer network that is fine in modeling the interactive relationships among the temporal and frequency components of acoustic features by alternate blocks.
	\item We adopt self-supervised learning strategies to train the network with normal sounds. 
	These strategies adequately explore the information derived from the normal sounds. 
	By mapping the acoustic features with their machine identity information and jointly reconstructing the masked acoustic features, the SSDPT can learn discriminative and robust latent acoustic representations.
	\item To the best of our knowledge, we are the first to propose SSDPT to detect anomalous sound for machine condition monitoring. Experimental results show that our approach achieves state-of-the-art performance.
\end{itemize}

The rest of this paper is organized as follows: 
Section \ref{sec:proposed method} introduces the proposed SSDPT. 
Section \ref{sec:Experiments} describes the details of the experiments. 
Section \ref{sec:results} gives the results and discussion.
Section \ref{sec:conclusion} concludes this paper.

\begin{figure*}[h]
	\begin{center}
		\includegraphics[scale=0.7]{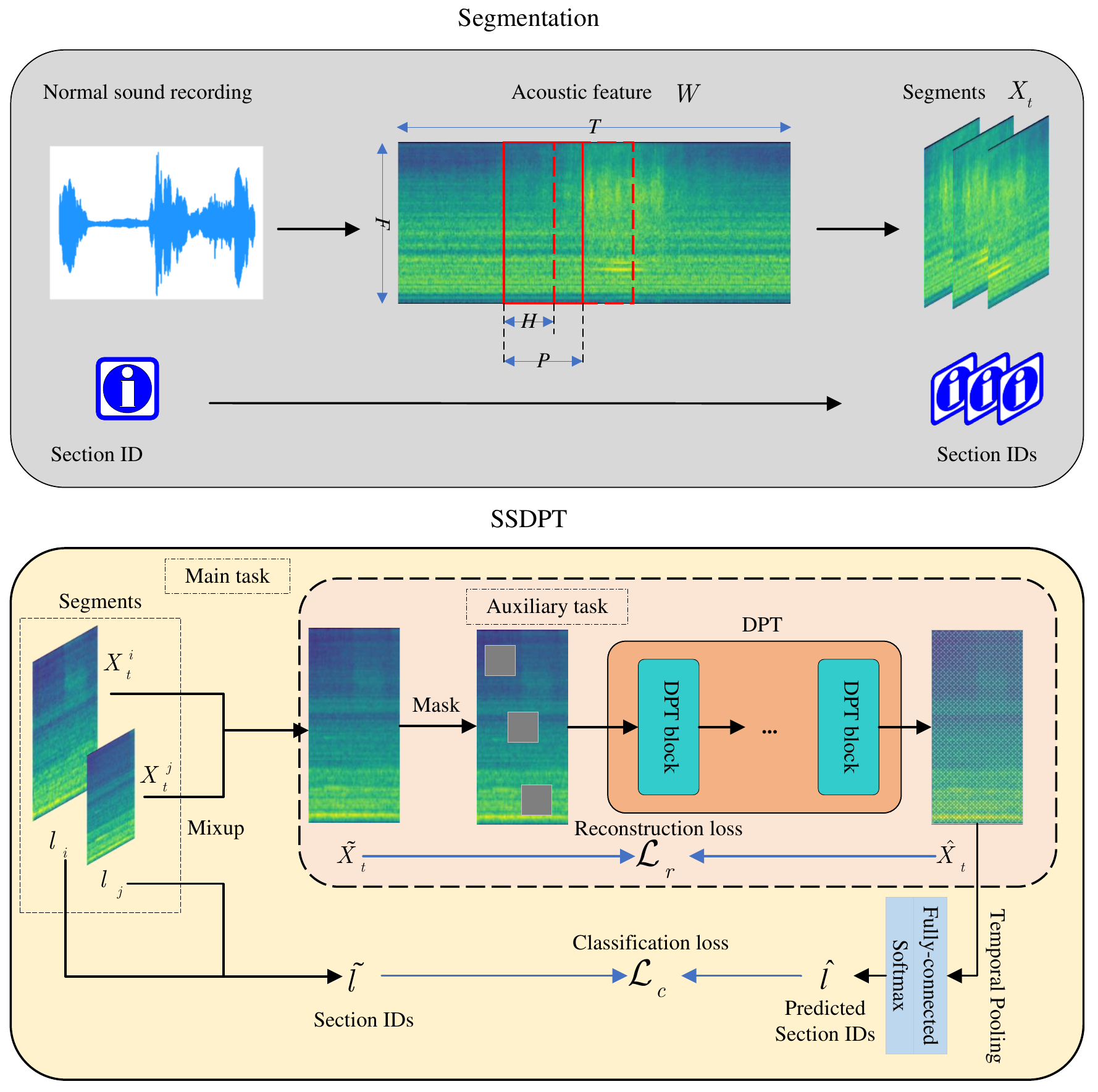}
	\end{center}
	\caption{The flowchart of the proposed SSDPT. }
	\label{fig:DPT_overview}
\end{figure*}

\section{Proposed Method}
\label{sec:proposed method}
In this section, we first introduce the architecture of a general ASD system.
We then introduce the proposed SSDPT for ASD. 
The flowchart of the proposed SSDPT is shown in Fig. \ref{fig:DPT_overview}.
Finally, we describe the modules of SSDPT in detail, including segmentation, DPT block, and SSDPT.

\subsection{ASD systems}
\label{subsec:ASD_systems}
Anomaly detection is usually regarded as an outlier detection problem since anomalous data is not available during training.
Therefore, ASD systems usually use unsupervised learning methods for anomaly detection.
A normal group of unsupervised learning methods consists of reconstruction-based approaches, for example, AEs.
The overall processing stages of an AE-based ASD system are described as follows.

In the training stage, a sample of normal sound $x\in\mathbb{R}^{L}$ is transformed into a time-frequency acoustic feature $X \in \mathbb{R}^{P\times F}$ of $P$ frames and $F$ frequency bins.
The AE takes the acoustic feature $X$ as input, which is encoded as a latent feature by the encoder and reconstructed by the decoder. 
The aim of AE is to minimize the reconstruction loss between the input and output:
\begin{equation}
\mathcal{L}_{AE}(X, \theta)=||X-\hat{X}||_{2}^{2}=||X- \mathcal{F}_{AE}(X, \theta)||_{2}^{2},
\end{equation}
where $\hat{X}$ is the reconstructed acoustic feature, $\mathcal{F}_{AE}(\cdot,\theta)$ is the AE model with training parameters $\theta$, and $\mathcal{L}_{AE}$ is the reconstruction loss of AE.

In the testing stage, an unsupervised ASD system has to identify whether a given test sound sample $x$ is anomalous.
The ASD system needs to calculate an anomaly score $\mathcal{A}(x)$.
It is assumed that in anomaly detection there is a bound of normal data.
$x$ is regarded as anomalous when the anomaly score is over a threshold $\tau$, otherwise, it is regarded as normal:
\begin{equation}
\mathcal{C}(x, \tau)=\left\{
\begin{aligned}
&0(Normal), \mathcal{A}(x)<\tau \\
&1 (Anomaly), \mathcal{A}(x)\ge\tau.
\end{aligned}
\right.
\end{equation}

\subsection{Segmentation}
\label{subsec:Segmentation}
SSDPT first transforms normal sound recordings into acoustic features using feature extraction.
We assume that each acoustic feature of one machine type is assigned with the ID $l$.
Then each acoustic feature $W\in \mathbb{R}^{T\times F}$ will be divided into overlapping short segments $X_{t}\in \mathbb{R}^{P\times F}, t=1,\dots,B$, where $B=\operatorname{int}(T/H)$, using a frame length of $P$ and a hop length of $H$.
Meanwhile, each short segment is assigned with the same machine ID $l$.

\subsection{DPT block}
The segmented acoustic features are taken as the input of the ASD model.
We propose DPT to alternately model the interrelations of temporal and frequency components of the acoustic features.
DPT consists of several stacked DPT blocks, where each block comprises two sub-modules, which are based on Transformer encoders.

Transformer is a sequence-to-sequence model. 
We utilize the Transformer encoders to sequentially model on two dimensions of the acoustic features.
The Transformer encoder used in DPT consists of MHSA and a position-wise feed-forward network (FFN).
In the MHSA module, multiple scaled dot product attention modules are first applied to the acoustic features.
The attention of all heads is linearly concatenated and computed on the elements of feature sequences. 
We use residual connections and layer normalization (LN) \cite{ba2016layer} on the output of MHSA, and feed the output of MHSA to the FFN with residual connections and LN.
These processing steps are formulated as:
\begin{equation}
X_{MHSA} = MHSA\left(X_{t} \right),
\end{equation}
\begin{equation}
X_{Mid} = LN\left( X_{MHSA}+X_{t}  \right),
\end{equation}
\begin{equation}
X_{FFN} = FFN\left(X_{Mid} \right),
\end{equation}
\begin{equation}
X_{O} = LN\left( X_{FFN}+X_{Mid}  \right),
\end{equation}
where $X_{t}$ is the input acoustic feature, $X_{MHSA}$ is the output of MHSA, $X_{FFN}$ is the output of FFN, and $X_{O}$ is the output of the Transformer encoder.

\begin{figure}[t]
	\centering
	\includegraphics[scale=0.8]{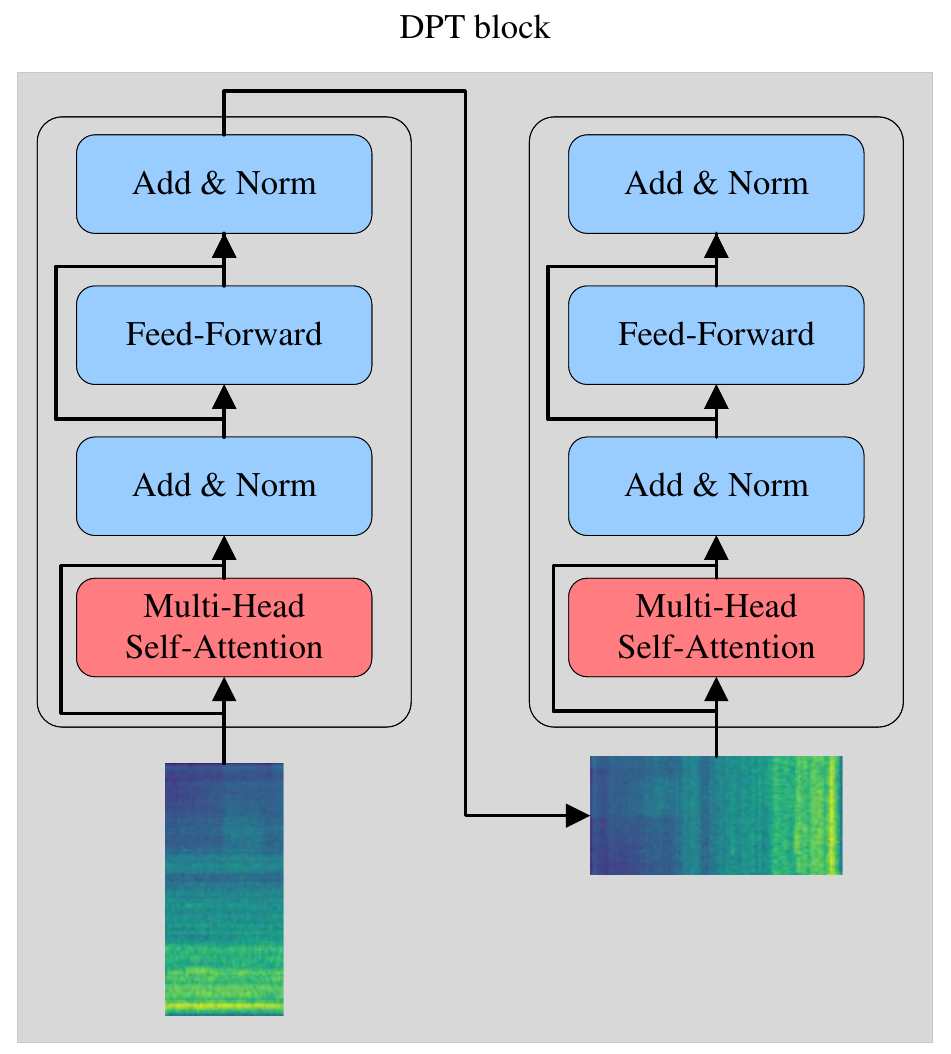}
	\caption{Architecture of a DPT block.}
	\label{fig:DPT_block}
\end{figure}

We present the architecture of a DPT block in Fig. \ref{fig:DPT_block}.
In the DPT block, $X_{t}$ is taken as the input of the first Transformer encoder and modeled on the time dimension:
\begin{equation}
X^{s}_{t} = \mathcal{E}_{s}\left(X_{t} \right),
\end{equation}
where $X^{s}_{t}\in \mathbb{R}^{P\times F}$ is the output of the first  Transformer encoder, and $\mathcal{E}_{s}(\cdot)$ denotes the mapping function of the first Transformer encoder.
$X^{s}_{t}$ is then transposed and provided as the input of the second Transformer encoder, which models on the frequency dimension:
\begin{equation}
X^{f}_{t} = \mathcal{E}_{f}\left((X^{s}_{t})^{T} \right),
\end{equation}
where $X^{f}_{t}\in \mathbb{R}^{F\times P}$ is the output of the second Transformer encoder, and $\mathcal{E}_{f}(\cdot)$ denotes the mapping function of the second Transformer encoder.
Then, $X^{f}_{t}$ will be transposed and passed to the next DPT block, where the number of blocks in the proposed DPT is denoted as $M$.

\subsection{SSDPT}

We combine two SSL strategies in our SSDPT network to improve the performance of ASD.
The first one is based on a classification approach and serves as the main task of SSDPT. 
This approach aims to train the model using normal sounds with different machine IDs for the same machine type.
The model tends to learn robust representations of normal sound, and such representations are inherent parts of the machines, e.g., size or manufacturer.
The second approach is a subcategory of reconstruction methods \cite{li2020superpixel, ristea2021self}, serving as the auxiliary task in SSDPT.
This approach relies on predicting masked information of the input acoustic features with different forms of masks.
By introducing the auxiliary task, the proposed SSDPT can potentially learn more discriminative time-frequency structure of the acoustic features.
Moreover, we employ mixup \cite{zhang2017mixup} as the data augmentation method to improve the generalization of the model.

In the training stage, each input feature $X_{t}$ is assigned with a machine ID $l\in \{1, \dots, I\}$, where $I$ is the number of IDs.
The operations of mixup are expressed as follows:
\begin{equation}
\centering
\tilde{X}_{t}=\lambda X^{i}_{t}+(1-\lambda) X^{j}_{t},
\end{equation}
\begin{equation}
\tilde{l}=\lambda l_{i}+(1-\lambda) l_{j},
\end{equation}
where $\tilde{X}_{t}$ is the mixed acoustic feature, $\tilde{l}$ is the mixed machine ID, $X^{i}_{t}$ and $X^{j}_{t}$ are two acoustic features, the machine IDs are represented by $l_{i}$ and $l_{j}$, and $\lambda\in[0,1]$ represents a random number that is taken from a beta distribution.
We then generate $k$ masks and apply them to the mixed acoustic feature $\tilde{X}_{t}$.
After that, the mixed and masked acoustic feature is fed to the proposed DPT.
% loss-c loss-r loss-all
Assuming $\hat{X}_{t}$ is the output of the DPT, we calculate the max values along the $P$ dimension of $\hat{X}_{t}$, and apply a fully-connected (FC) layer along with a softmax function to get the probability vector $z_{t} \in \mathbb{R}^{I}$.
% \begin{equation}
% z_{t}= softmax\left(FC\left( X^{\prime}_{t}\right)\right),
% \end{equation}
The predicted label $\hat{l}$ is the index of the max value of $z_{t}$.
We use cross-entropy as the classification loss $\mathcal{L}_{c}$ for the main task, which is formulated as follows:
\begin{equation}
\mathcal{L}_{c} = CrossEntropy\left( \hat{l}, \tilde{l} \right),
\end{equation}
and we use mean square error (MSE) as the reconstruction loss $\mathcal{L}_{r}$ for the auxiliary task, formulated as:
\begin{equation}
\mathcal{L}_{r} = || \tilde{X}_{t} - \hat{X}_{t}||^{2}_{2}.
\end{equation}
The SSDPT is trained by jointly minimizing the classification error $\mathcal{L}_{c}$ and the reconstruction error $\mathcal{L}_{r}$.
The overall training object $\mathcal{L}$ is given as follows:
\begin{equation}
\mathcal{L} = \frac{1}{N}\sum^{N}_{n=1}(\mathcal{L}_{c}+\alpha \mathcal{L}_{r}),
\end{equation}
where $\alpha > 0$ is a hyper-parameter, and $N$ is the number of training samples of normal sounds. 

% score-c score-r score-all
% Anomaly score b*Lr+logLc
In the testing stage, we will calculate an anomaly score for a given test sound to determine if it is anomalous.
We assume that the acoustic feature of the test sample is $W^{\prime}\in \mathbb{R}^{T\times F}$ then divide it into $B$ overlapping segments $X^{\prime}_{t}\in \mathbb{R}^{P\times F}$.
The anomaly score of the main task is calculated as:
\begin{equation}
\mathcal{A}_{c}(z^{\prime}_{t}) = \frac{1}{B} \sum_{b = 1}^B \log {\LARGE \{} \frac{1 - p(z^{\prime}_{t})}{p(z^{\prime}_{t})} {\LARGE \}},
\end{equation}
where $z^{\prime}_{t}$ is the probability vector, and $p(\cdot)$ is the output for the corresponding ID.
The anomaly score of the auxiliary task is the MSE between  $X^{\prime}_{t}$ and its reconstruction feature  $\hat{X}^{\prime}_{t}$:
\begin{equation}
\mathcal{A}_{r}(X^{\prime}_{t}) = \frac{1}{B} \sum_{b = 1}^B || X^{\prime}_{t} - \hat{X}^{\prime}_{t}||^{2}_{2}.
\end{equation}
Finally, we calculate the total anomaly score $\mathcal{A}$ as:
\begin{equation}
\mathcal{A} = \mathcal{A}_{c}+\beta \mathcal{A}_{r},
\end{equation}
where $\beta > 0$ is a hyper-parameter.

The learning procedure of the SSDPT is summarized in Algorithm \ref{alg:SSDPT}.
\begin{algorithm}[h!]
\caption{Learning Algorithm of the Proposed Self-Supervised Dual-Path Transformer}
	\label{alg:SSDPT}
	\begin{algorithmic}[1]
		\REQUIRE Training acoustic features $\mathcal{D}_{tr}=(W_{s},l_{s})_{s=1}^{S_{1}}$;\\
		Test acoustic features $\mathcal{D}_{te}=(W^{\prime}_{s},l_{s})_{s=1}^{S_{2}}$;\\
		DPT model $\mathcal{F}_{DPT}(\cdot, \theta)$;\\
		Hyper-parameters $\alpha$ and $\beta$;
		\ENSURE  Anomalous scores $(\mathcal{A}(W^{\prime}_{s}))_{s=1}^{S_{2}}$.
		\STATE {\textbf{Segmentation} Divide $W_{t}$ into overlapped  short segments $X_{t}$ with machine IDs $l$}
		\STATE {Initialize parameters $\theta$, $\alpha>0$ and $\beta>0$}
		\FOR{each epoch}
		\FOR{$n=1,...,N$}
		\STATE{Sample two acoustic features $X^{i}_{t}$ and $X^{j}_{t}$ with IDs $l_{i}$ and $l_{j}$\\
			\textbf{Mixup} $\tilde{X}_{t}$ $\gets$ Mix $X^{i}_{t}$ and $X^{j}_{t}$,$\quad\tilde{l}$ $\gets$ Mix $l_{i}$ and $l_{j}$\\
			\textbf{Mask} $\tilde{X}_{t}$\\
			$\hat{X}_{t},\ \hat{l}$ $\gets$ $\mathcal{F}_{DPT}(\tilde{X}_{t}, \theta)$\\                $\mathcal{L}_{c}= CrossEntropy\left( \hat{l}, \tilde{l} \right)$\\ 
    		$\mathcal{L}_{r}$ = $||\left( \tilde{X}_{t} - \hat{X}_{t}\right)||^{2}_{2}$\\
			$\mathcal{L} += \mathcal{L}_{c}+\alpha \mathcal{L}_{r}$\\
		}
		\ENDFOR	
		\STATE{
			$\mathcal{L} = \mathcal{L}/N$\\
			\textbf{update} $\theta$ in $\mathcal{F}_{DPT}(\cdot)$ to minimize $\mathcal{L}$}\\
		\ENDFOR
		\FOR{each $W^{\prime}_{s}$ in $\mathcal{D}_{te}$}
		\STATE{Divide $W^{\prime}_{s}$ into $B$ overlapped short segments $X^{\prime}_{t}$}
		\FOR{$b=1,...,B$}
		\STATE{ 
			$\hat{X}_{t},\ z^{\prime}_{t}$ $\gets$ $\mathcal{F}_{DPT}(X^{\prime}_{t},\theta)$\\
			\textbf{Calculate} $\mathcal{A}_{c}(z^{\prime}_{t})$
			$\mathcal{A}_{r}(X^{\prime}_{t})$ }
		\ENDFOR
		\STATE{ 
			\textbf{Calculate} $\mathcal{A} = (\mathcal{A}_{c}+\beta \mathcal{A}_{r})/B$
		}
		\ENDFOR
\end{algorithmic}
\end{algorithm}

\section{Experiments}
\label{sec:Experiments}
\subsection{Dataset}
We use the development dataset of the Detection and Classification of Acoustic Scenes and Events (DCASE) 2021 task2 for evaluating the proposed SSDPT.
This dataset consists of sounds from seven machine types, including toyCar, toyTrain, fan, gearbox, pump, slider rail, and valve \cite{Tanabe_arXiv2021_01, harada2021toyadmos2}.
Each clip is 10-second audio with a 16,000 Hz sampling rate, including operating sound and background noise. 
The development dataset contains three sections as machine IDs for each machine type, i.e., "Section 00", "Section 01", and "Section 02".
In each section, the data are divided into two different conditions, source and target domain. 
For the training, there are 1,000 normal clips for the source domain and 3 normal clips for the target domain.
For the test, there are 100 normal and 100 anomalous clips for both the source and target domains.

\subsection{Experimental Setups}
\label{subsec:Experimental Setups}
\begin{figure}
	\centering
	\includegraphics[scale=1]{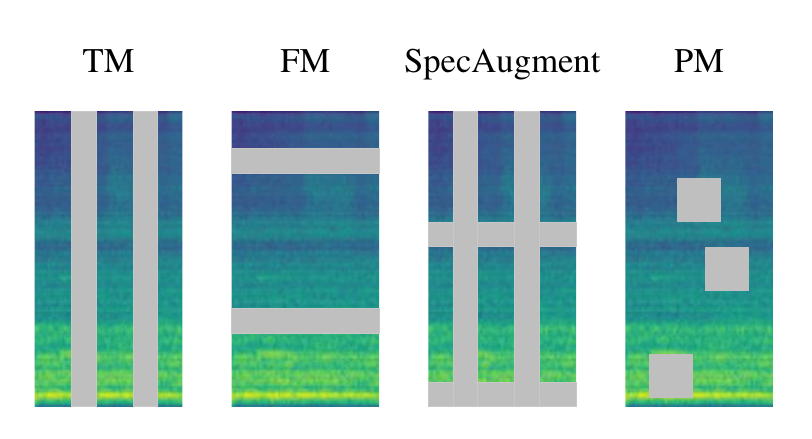}
	\caption{Examples of acoustic features masked by different methods.
		Time masking (TM), frequency masking (FM), and patch masking (PM).}
	\label{fig:Masking}
\end{figure}

% frames/blocks/mask methods/cutout number&size/ score weight
The input acoustic features of the proposed method are log-Mel spectrograms.
We used the short-time Fourier transform with a window size of 1024 and hop length of 512 to generate spectrograms from the audio signals.
Mel filters of 128 bands ($F$) are used to transform the spectrograms into Mel spectrograms. 
We applied logarithm on Mel spectrograms to get the log-Mel spectrograms.
In the segmentation stage, we experimented with different frame lengths $P$ of 64, 128, and 256 with a hop length $H$ of 8 to segment the acoustic features of normal sounds. 
We set the hop length to 1 for generating segmented acoustic features for test sounds.
Moreover, we introduced different methods to mask the areas of the acoustic features, including time masking (TM), frequency masking (FM), SpecAugment, and patch masking (PM).
Examples of masked acoustic features with different methods are shown in Fig. \ref{fig:Masking}.

We experimented with different numbers ($M=1, 2, 3$) of DPT blocks, where the Transformer encoders use the same configuration of 1 encoder layer, 8 MHSA heads, and 32 FFN nodes.
Especially, in each DPT block, the embedding size and the sequence length of $\mathcal{E}_{t}(\cdot)$ is equal to $P$ and $F$, while for $\mathcal{E}_{f}(\cdot)$ the embedding size and the sequence length are $F$ and $P$, respectively.

We used the normal sound recordings from the source and target domain to train SSDPT. AdamW \cite{loshchilov2018fixing} is adopted as the optimizer, and we used a dynamic strategy to adjust the learning rate during training with an initial learning rate of 0.0001.
The default values of the hyper-parameter $\alpha$ and $\beta$ are set to 0.001.

\subsection{Baseline systems}

To evaluate the performance of SSDPT for ASD, we compare the proposed SSDPT with the following methods:

\textbf{Official baselines \cite{kawaguchi2021description} }
The organizers of the DCASE 2021 task2 provided two baseline systems. 
The first approach is an AE-based system, in which 5 consecutive frames of log-Mel spectrogram with bands of 128 are concatenated and taken as the input vector. 
The training of the baseline is done to reduce the MSE of normal sounds, and the mean reconstruction error is based on the anomaly score.
The second approach is a classification-based baseline using the machine IDs with MobileNetV2 (MNv2) \cite{sandler2018mobilenetv2}. 
This baseline takes 64 consecutive frames of log-Mel spectrogram with bands of 128 as input acoustic feature, and the anomaly score is calculated as the averaged negative logarithm of predicted probabilities for the corresponding ID.

\textbf{WavNet-Ensemble \cite{lopez2021ensemble}} This approach is proposed by the top1 challengers of the DCASE task2. 
They adopted a WaveNet \cite{oord2016wavenet} model with an x-vector component and AMS for directly processing the normal sound signals instead of acoustic features.
They also used different reprocessing and parameters for different machine types to improve the performance of ASD.

\textbf{MNv2-LOF \cite{morita2021anomalous}} This approach is proposed by the top2 challengers of the DCASE task2.
The challengers use a classification-based system with MNv2 and additive angular margin loss. 
This method takes spectrograms with 1024 dimensions and 32 frames as input acoustic features. 
Moreover, they adopted the local outlier factor (LOF) \cite{breunig2000lof} as the anomaly detector, where the output of MNv2 is used as the input of LOF.

\subsection{Evaluation metrics}
The evaluate the proposed SSDPT, we use the area under the receiver operating characteristic curve (AUC) and the partial AUC (pAUC) \cite{kawaguchi2021description} as the evaluation metrics. 
The AUC score for each machine type, section, and domain is calculated as:
\begin{equation}
\text{AUC} = \frac{1}{N_{-}N_{+}} \sum_{i=1}^{N_{-}} \sum_{j=1}^{N_{+}} \mathcal{C} (\mathcal{A}^{+} - \mathcal{A}^{-}),
\end{equation}
where $\mathcal{C}(x)$ returns 1 when $x>0$ and 0 otherwise, $\mathcal{A}^{+}$ and $\mathcal{A}^{-}$ are anomaly scores of normal and anomalous sounds, $N_{+}$ and $N_{-}$ are the number of normal and anomalous sounds, respectively.
We assume that $\mathcal{A}$ follows a gamma distribution, and we determine the anomaly detection threshold of pAUC under the condition that the false positive rate is 0.1, which is equal to $p$.
The pAUC score for each machine type, section, and domain is calculated as:
\begin{equation}
\text{pAUC} = \frac{1}{\left \lfloor pN_{-} \right \rfloor N_{+}} \sum_{i=1}^{\left \lfloor pN_{-} \right \rfloor} \sum_{j=1}^{N_{+}} \mathcal{C} (\mathcal{A}^{+} - \mathcal{A}^{-}),
\end{equation}
where $\left \lfloor \cdot \right \rfloor$ is the flooring function.
The harmonic mean values of AUC (h-AUC) and pAUC (h-pAUC) over sections and domains for each machine type are calculated for the performance comparison of ASD.

\section{Results and Discussion}
\label{sec:results}
In this section, we introduce the experimental results and discussions based on two aspects.
We first present the overall results and compare the proposed SSDPT against the baseline systems.
We then do ablation experiments of the SSDPT for further analysis.

\subsection{Overall results}
% baselines vs DPT vs SSDPT vs SSDPT-ensemble 
% 柱状图每一类
\begin{table}[h]
\centering
\renewcommand\arraystretch{1.25}	
\caption{Comparison with other methods in terms of h-AUC and h-pAUC.}
\label{Overall results}
\begin{tabular}{cccccccclcl}
    \hline\hline
    Method           &  &  &  &  &  &  & \multicolumn{2}{c}{h-AUC} & \multicolumn{2}{c}{h-pAUC} \\ \hline
    Baseline-AE \cite{kawaguchi2021description}      &  &  &  &  &  &  & \multicolumn{2}{c}{0.619} & \multicolumn{2}{c}{0.533}  \\
    Baseline-MNv2 \cite{kawaguchi2021description}   &  &  &  &  &  &  & \multicolumn{2}{c}{0.597} & \multicolumn{2}{c}{0.564}  \\
    MNv2-LOF \cite{morita2021anomalous}       &  &  &  &  &  &  & \multicolumn{2}{c}{0.694} & \multicolumn{2}{c}{0.616}  \\
    WaveNet-ensemble \cite{lopez2021ensemble}  &  &  &  &  &  &  & \multicolumn{2}{c}{0.705} & \multicolumn{2}{c}{0.625}  \\ \hline
    SSDPT(w/o auxiliary task)   &  &  &  &  &  &  & \multicolumn{2}{c}{0.713} & \multicolumn{2}{c}{0.607}  \\
    SSDPT            &  &  &  &  &  &  & \multicolumn{2}{c}{0.722}  & \multicolumn{2}{c}{0.611}  \\
    SSDPT-ensemble   &  &  &  &  &  &  & \multicolumn{2}{c}{\textbf{0.739}}      & \multicolumn{2}{c}{\textbf{0.626}}       \\ \hline\hline
\end{tabular}
\end{table}

Table \ref{Overall results} shows the h-AUC and h-pAUC of the proposed SSDPT compared with the baseline systems on the development dataset of DCASE 2021 task2.
MNv2-LOF and WaveNet-ensemble achieve state-of-the-art performance in task2, significantly surpassing the two official baseline systems, i.e., Baseline-AE and Baseline-MNv2.
The proposed SSDPT networks improve the performance compared to the MNv2-LOF and WaveNet-ensemble.
Overall, the results of our SSDPT networks are better than those of the comparative baseline systems. 
Specifically, SSDPT outperforms the state-of-the-art systems and achieves an h-AUC and h-pAUC score of 0.722 and 0.611, respectively.
To further improve the performance, we made an ensemble of the SSDPT models with different configurations, the SSDPT-ensemble significantly boosts the h-AUC and h-pAUC to 0.739 and 0.626, respectively.

\begin{figure}[t]
	\centering
	\includegraphics[scale=0.5]{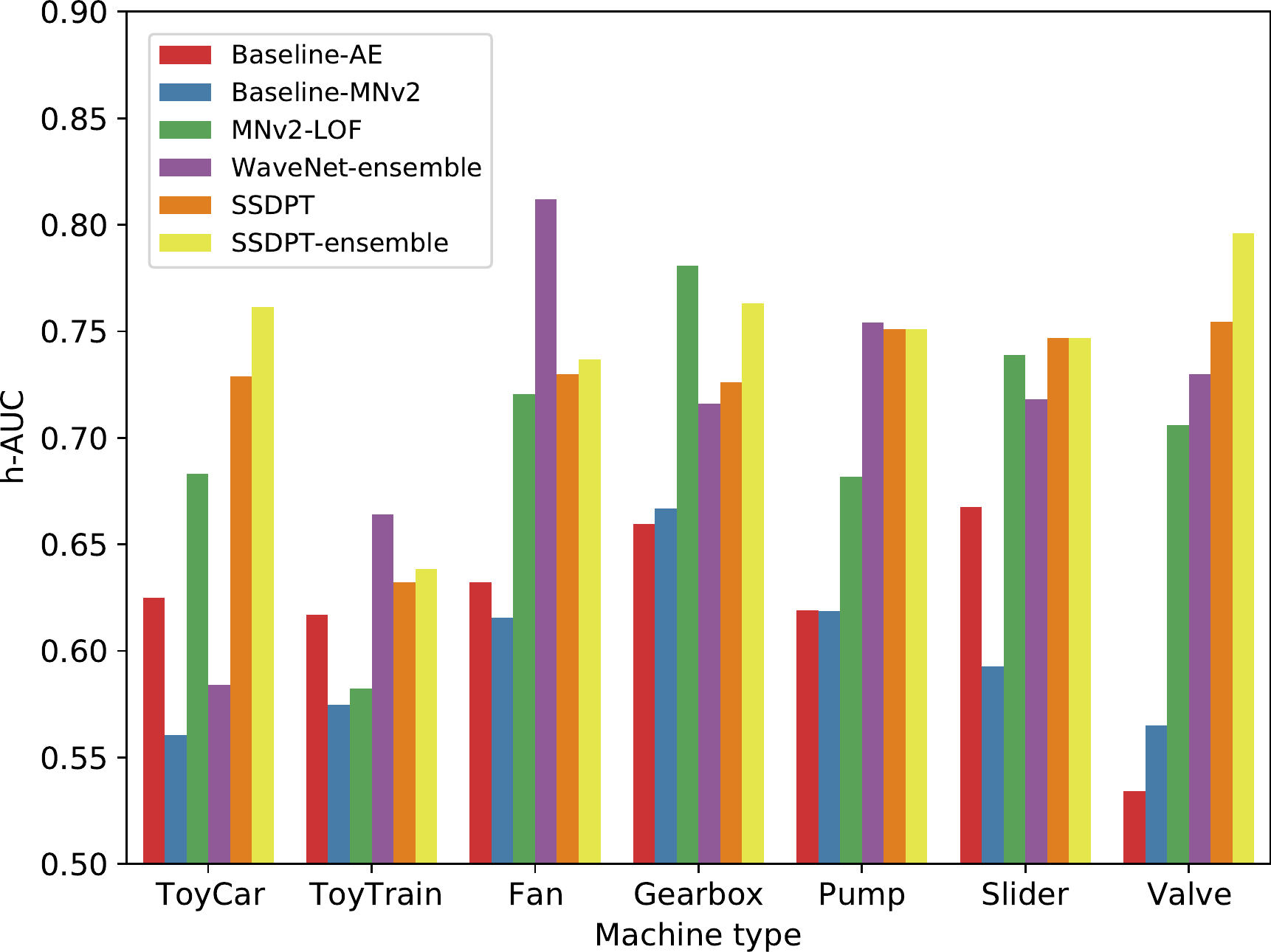}
	\caption{Results of comparative methods over different machine types.}
	\label{fig:Methods_results}
\end{figure}

Fig. \ref{fig:Methods_results} shows the h-AUC scores of the proposed SSDPT networks compared with baseline systems over different machine types.
WaveNet-ensemble achieves better results for toyTrain and fan, and MNv2-LOF achieves better results for gearbox.
The proposed SSDPT-based methods outperform the baseline systems for toyCar, slider, and valve.
Specifically, SSDPT-ensemble significantly improves the performance of SSDPT for toyCar, gearbox, and valve.

Moreover, we adopted two SSL methods to train the network in the SSDPT.
The classification-based method is used as the main task, while the reconstruction-based method is used as the auxiliary task.
We also experimented SSDPT without introducing the reconstruction-based auxiliary task for ASD.
In Table \ref{Overall results}, the results show that SSDPT without auxiliary task achieves an h-AUC and h-pAUC score of 0.713 and 0.607, respectively.
We then introduced the auxiliary task to the SSDPT. 
The h-AUC and h-pAUC is improved to 0.722 and 0.611, respectively.
This indicates that introducing the reconstruction-based auxiliary task can improve the performance for ASD.

\subsection{Ablation results}
% mask methods/cutout number&size/frames/blocks/score weight
The ablation experiments are demonstrated in this section, including masking methods, configurations of the DPT architecture, and hyper-parameter $\beta$ for calculating anomaly score.

\subsubsection{Masking methods}
% masking method 示意图
% NM TM FM PM柱状图 + PM 尺寸 柱状图
In Sec \ref{subsec:Experimental Setups}, we showed examples of different masking methods: TM, FM, SpecAugment, and PM. 
SpecAugment \cite{park2019specaugment} is a simple but effective masking method, which has been applied to many environmental sound recognition tasks.
It randomly masks the frequency bins and time frames of the acoustic features.
TM and FM only cover time frames or frequency bins.
While PM randomly covers $k$ square areas of the acoustic features.

\begin{figure}[t]
	\centering
	\includegraphics[scale=0.5]{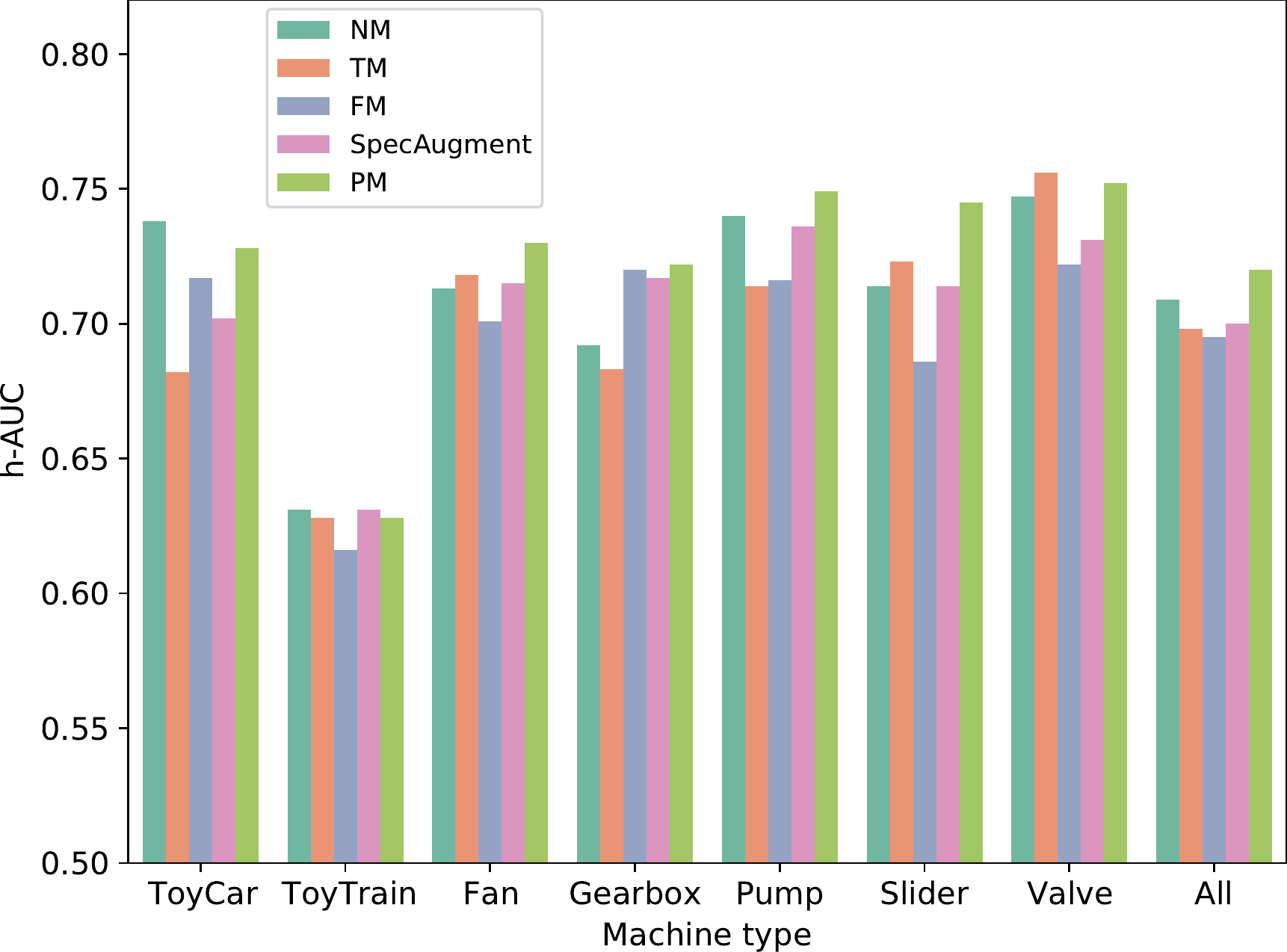}
	\caption{Results of masking methods.
		No masking (NM), time masking (TM), frequency masking (FM), and patch masking (PM).}
	\label{fig:Masking_method}
\end{figure}

We first experimented with these masking methods in SSDPT.
TM applies 2 masks with 4 consecutive frames, FM applies 2 masks with 4 consecutive frequency bins, SpecAugment applies the masks from both TM and FM, and PM applies 3 square masks each with a size of $5*5$. 
Fig. \ref{fig:Masking_method} shows the results of different masking methods over machine types.
PM outperforms the other methods over all machines, and achieves the best performance for fan, gearbox, pump, and slider.
NM achieves the best performance for toyCar and toyTrain.
However, the performance of applying TM, FM, and SpecAugment is not as good as supposed.
We assume that TM, FM, and SpecAugment mask the important time-frequency information of the acoustic features for some machine types, leading to a decrease in the performance.
While the proposed SSDPT can learn salient and robust time-frequency features by applying PM.

\begin{figure}[t]
	\centering
	\includegraphics[scale=0.5]{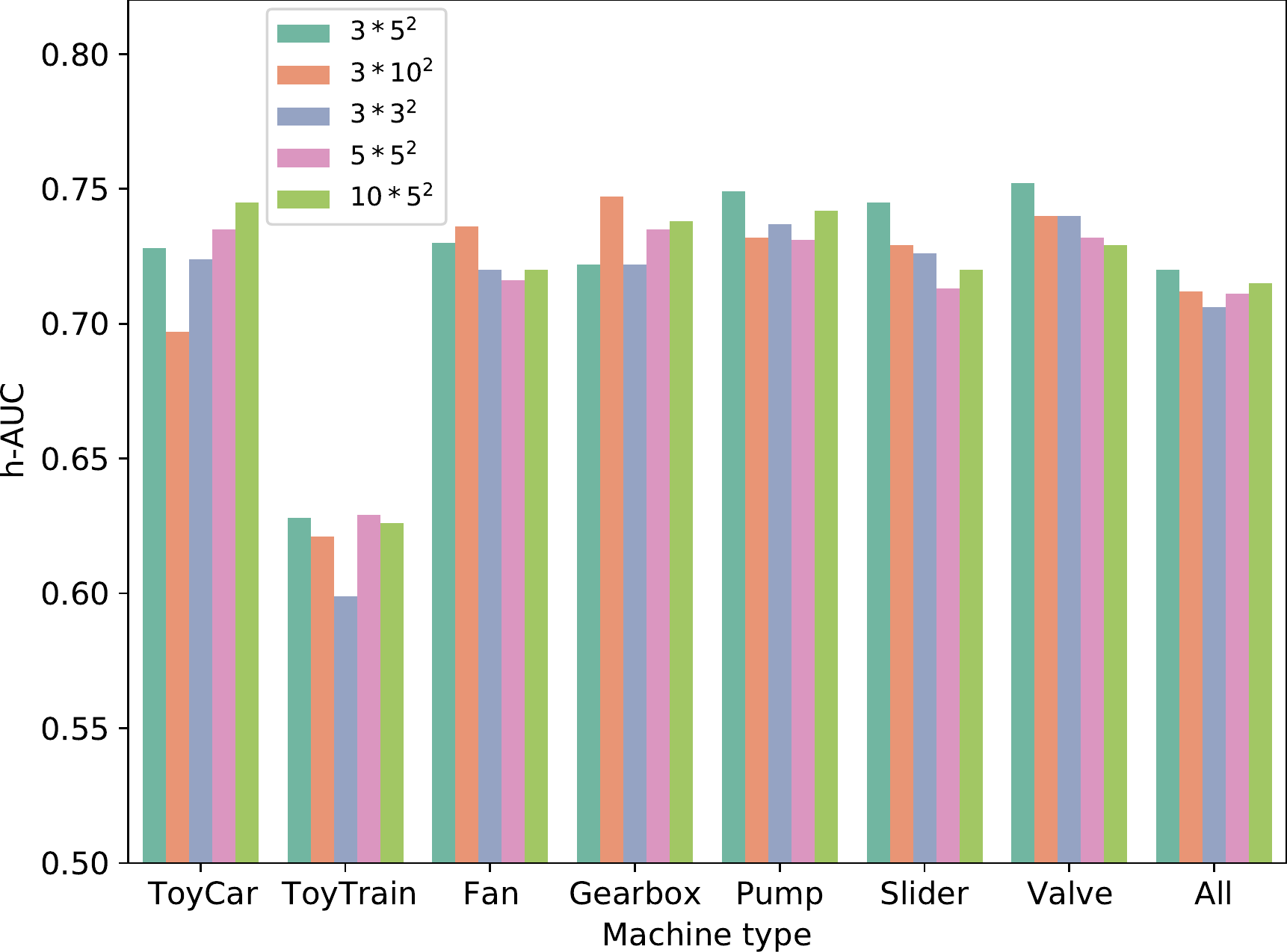}
	\caption{Results of PM with different numbers and sizes of masks.
		$k*r^{2}$ means $k$ square masks with a size of $r*r$, e.g., $3*5^{2}$ means 3 square masks with a size of $5*5$.}
	\label{fig:PM_results}
\end{figure}

We then experimented with PM using different mask numbers and sizes. 
Experimental results are shown in Fig. \ref{fig:PM_results}.
PM with $3*5^{2}$ outperforms the other settings of masks in total, and achieves the best performance for pump, slider, and valve.
PM with $3*10^{2}$ performs the best for fan and gearbox.
From Fig. \ref{fig:PM_results}, it is observed that different PM methods achieve different performance, the choice of number and size of the mask is related to the characteristics of the machine type.

\subsubsection{DPT Configurations}
% 表格
% \usepackage{multirow}

\begin{table*}[h]
	\centering
	\caption{Results of different configurations of DPT in terms of h-AUC.}
	\label{tab:Results of settings}
        \renewcommand\arraystretch{1.25}
	\resizebox{\textwidth}{!}{
		\begin{tabular}{ccccccccccc}
			\hline\hline
			Blocks ($M$) & Frame length ($P$) & ToyCar  & ToyTrain  & Fan   & Gearbox        & Pump           & Slider         & Valve          & All   & Parameters    \\ \hline
			1          & 64               & 0.686          & 0.602          & 0.721          & 0.687          & 0.696          & 0.726          & 0.712          & 0.688 & \textbf{13,955} \\
			2          & 64               & 0.702          & 0.626          & \textbf{0.732} & \textbf{0.759} & 0.713          & 0.728          & 0.754          & 0.714 & 27,267         \\
			3          & 64               & 0.728          & 0.628          & 0.73           & 0.722          & \textbf{0.749} & \textbf{0.745} & 0.752          & \textbf{0.720} & 40,579          \\
			3          & 128              & \textbf{0.758} & \textbf{0.638} & 0.728          & 0.696          & 0.715          & 0.636          & \textbf{0.794} & 0.705 & 53,827          \\
			3          & 256              & 0.706          & 0.602          & 0.699          & 0.695          & 0.719          & 0.671          & 0.782          & 0.693 & 80,323         \\ \hline\hline
	\end{tabular}}
\end{table*}

Table \ref{tab:Results of settings} compares the performance of SSDPT with different frame lengths and the numbers of DPT blocks.
We observe that the performance of toyCar, toyTrain, and valve is significantly improved by SSDPT with a frame length of 128. 
While longer frame length does not perform well for all machine types, indicating that the frame length is related to the inherent characteristics of different machine types.
Moreover, we found that increasing the number of DPT blocks leads to better performance of ASD, while DPT with 2 blocks achieves better results for gearbox.
Considering the number of parameters, DPT with 1 block can achieve a comparable h-AUC score of 0.688 with very few parameters, this facilitates the implementation of the proposed SSDPT in ASD applications.

\subsubsection{Hyper-parameter of the anomaly score}
\begin{figure}
	\centering
	\includegraphics[scale=0.45]{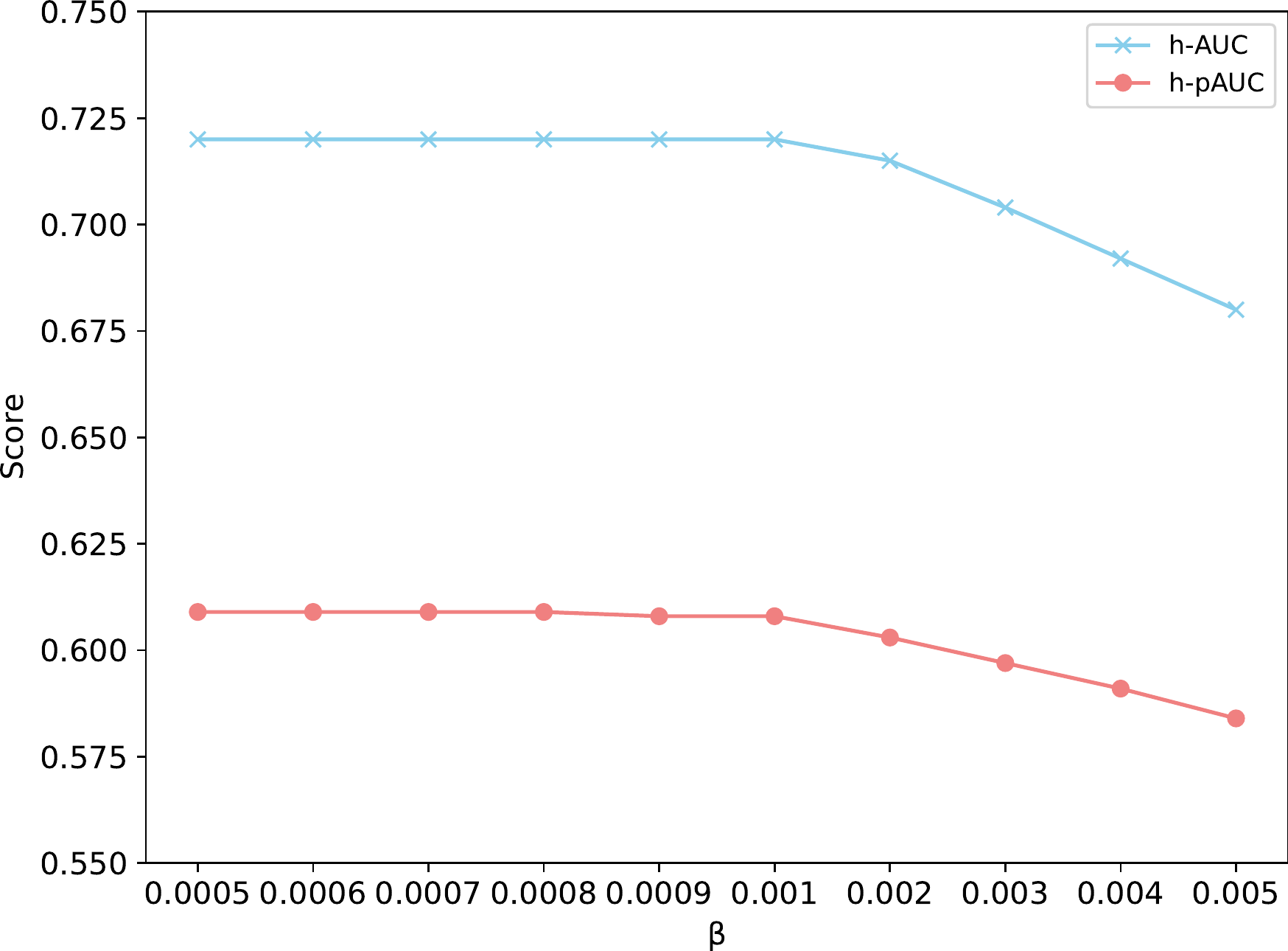}
	\caption{Performance comparison with different values of $\beta$ in terms of h-AUC and h-pAUC.}
	\label{fig:AUC_vs_weights}
\end{figure}

% 曲线+箱图
Previous experiments were conducted under the default value of the hyper-parameter $\beta$.
To analyze if the performance and anomaly score $\mathcal{A}$ will be affected by using different values of $\beta$, we changed $\beta$ from 0.0005 to 0.005 with non-linear intervals.
The experimental results are shown in Fig. \ref{fig:AUC_vs_weights}.
It can be seen that from 0.0005 to 0.001, the h-AUC and h-pAUC scores are relatively stable.
However, the h-AUC and h-pAUC scores start to decline when $\beta$ increases over 0.001.
Therefore, we suggest constraining $\beta$ between 0.0005 and 0.001.

\begin{figure}
	\centering
	\includegraphics[scale=0.5]{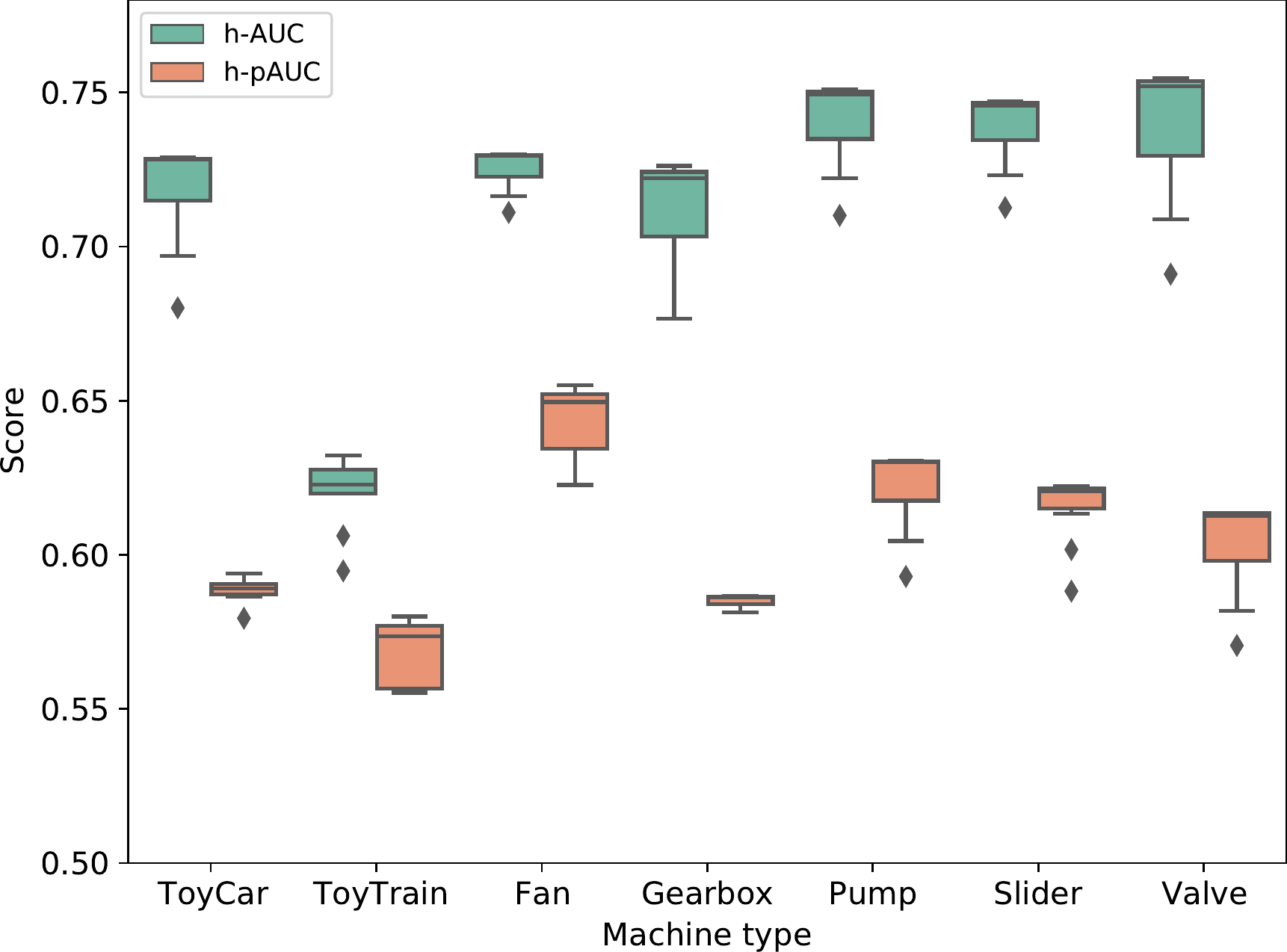}
	\caption{Performance comparison over each machine type with different values of $\beta$ in terms of h-AUC and h-pAUC.}
	\label{fig:Machine_vs_weights}
\end{figure}

Fig. \ref{fig:Machine_vs_weights} shows the boxplot using different values of $\beta$ for calculating h-AUC and h-pAUC over each machine type.
The height of the box indicates the dispersion of target data, and a larger height means that the data is more dispersed.
Overall, we can observe that the h-AUC and h-pAUC scores have different susceptibilities over machine types.
In detail, h-AUC scores of gearbox and valve are more susceptible to the values of $\beta$, while for h-pAUC, the scores of toyCar, gearbox, and slider are relatively stable and less influenced by different values of $\beta$.
This demonstrates that $\beta$ has to be adjusted carefully for various machine types.

\section{Conclusions}
\label{sec:conclusion}

In this paper, we proposed a self-supervised dual-path Transformer (SSDPT) based anomalous sound detection (ASD) framework for machine condition monitoring (MCM). 
In the SSDPT, log-Mel spectrograms of normal sounds are segmented into overlapped fragments and modeled using DPT. 
The proposed DPT consists of several alternate DPT blocks, in which the attention-based Transformer encoders are used for thoroughly modeling the temporal and frequency components of the acoustic features.
To further improve the performance of ASD, we adopted two self-supervised learning (SSL) strategies: the classification-based approach, which is severed as the main task for ASD using the machine section IDs for classification, and the reconstruction-based approach, which is introduced as the auxiliary task for ASD by masking the areas of the acoustic features and reconstructing them.

Experimental results show that the presented DPT outperforms state-of-the-art methods for ASD.
Moreover, the adoption of SSL strategies leads to better results on the evaluation metrics. 
We conclude that our proposed SSDPT adequately models the characteristics of the normal sounds and learns more robust acoustic representations.
We significantly improve the performance of ASD by the ensemble of SSDPT models, our approach achieves an h-AUC and h-pAUC score of 0.739 and 0.626, respectively.

%\section*{ACKNOWLEDGMENT}


\ifCLASSOPTIONcaptionsoff
\newpage
\fi



\bibliographystyle{IEEEtran}
\bibliography{bare_jrnl}
\end{document}